\documentclass[twocolumn,preprintnumbers,amsmath,amssymb,prl]{revtex4}

\usepackage{xspace}
\usepackage{graphicx}% Include figure files
\usepackage{dcolumn}% Align table columns on decimal point
\usepackage{bm}% bold math
\usepackage{tabularx}% Table width
\usepackage{amsmath}% Higher math
\usepackage{color}%color
\usepackage{ulem}

\begin{document}

\title{Heat transport of the kagom\'{e} Heisenberg quantum spin liquid candidate YCu$_3$(OH)$_{6.5}$Br$_{2.5}$: localized magnetic excitations and spin gap}

\author{Xi\v{a}och\'{e}n H\'{o}ng$^{1,2,}\footnote{xhong@uni-wuppertal.de}$, Mahdi Behnami$^2$, Long Yuan$^3$, Boqiang Li$^3$, Wolfram Brenig$^4$, Bernd B\"{u}chner$^{2,5}$, Yuesheng Li$^{3,}\footnote{yuesheng\_li@hust.edu.cn}$ and Christian Hess$^{1,2,}\footnote{c.hess@uni-wuppertal.de}$}

\affiliation{
$^1$Fakult$\ddot{a}$t f\"{u}r Mathematik und Naturwissenschaften, Bergische Universit$\ddot{a}$t Wuppertal, 42097 Wuppertal, Germany\\
$^2$Leibniz-Institute for Solid State and Materials Research (IFW-Dresden), 01069 Dresden, Germany\\
$^3$Wuhan National High Magnetic Field Center and School of Physics,Huazhong University of Science and Technology, 430074 Wuhan, China\\
$^4$Institute for Theoretical Physics, TU Braunschweig, 38106 Braunschweig, Germany\\
$^5$Institute of Solid State and Materials Physics and W\"{u}rzburg-Dresden Cluster of Excellence $ct.qmat$, Technische Universit$\ddot{a}$t Dresden, 01062 Dresden, Germany}

\date{\today}

\begin{abstract}
The spin-1/2 kagom\'{e} Heisenberg antiferromagnet is generally accepted as one of the most promising two-dimensional models to realize a quantum spin liquid state. Previous experimental efforts were almost exclusively on only one archetypal material, the herbertsmithite ZnCu$_3$(OH)$_6$Cl$_2$, which unfortunately suffers from the notorious orphan spins problem caused by magnetic disorders.
Here we turn to YCu$_3$(OH)$_{6.5}$Br$_{2.5}$, recently recognized as another host of a globally undistorted kagom\'{e} Cu$^{2+}$ lattice free from the orphan spins, thus a more feasible system for studying the intrinsic kagom\'{e} quantum spin liquid physics.
Our high-resolution low-temperature thermal conductivity measurements yield a vanishing small residual linear term of $\kappa/T$ ($T\rightarrow 0$), and thus clearly rule out itinerant gapless fermionic excitations. Unusual scattering of phonons grows exponentially with temperature, suggesting thermally activated phonon-spin scattering and hence a gapped magnetic excitation, consistent with a $\mathbb{Z}_2$ quantum spin liquid ground state. Additionally, the analysis of magnetic field impact on the thermal conductivity reveals a field closing of the spin gap, while the excitations remain localized.
\end{abstract}

\pacs{not needed}

\maketitle
The quantum spin liquid (QSL) is an intriguing state of matter that has fascinated physicists for decades. It is the ground state of certain magnetic systems where frustration prevents strongly correlated spins to form a well understood magnetically ordered ground state, but leave them as quantum paramagnetic/disordered states that are featured by long range topological ordering, fractionalized quasiparticles, and gauge excitations \cite {Balents,Knolle,Savary,ZhouY,Broholm}. Although appealing, our understanding of QSLs is still in its infancy. There is even no consensus whether QSL really exist in solid-state materials.  Apart from the bond dependent Kitaev proposals \cite{Kitaev}, strong geometric frustration is a prerequisite for QSL host materials. Natural candidates are those composed of magnetic ions forming triangle motifs \cite{JPCM}. In two dimensions, a highly frustrated geometry is the kagom\'{e} lattice \cite{Singh,Norman}, and the spin-1/2 kagom\'{e} Heisenberg antiferromagnet (the KHA model) is believed to hold great promise for implementing a QSL ground state.
Actually, herbertsmithite ZnCu$_3$(OH)$_6$Cl$_2$ has attracted enormous studies in the last decades as the archetypal realization of such model \cite{PALee,Shores,Helton,Norman,JPCS,HanTH,FuMX,HuangYY,Murayama,Khuntia}.

However, the ground state properties of herbertsmithite remain controversial \cite{Norman,JPCS,HanTH,FuMX,HuangYY,Murayama,Khuntia}.
Due to a similar chemical energy, an interlayer Cu/Zn antisite disorder is unavoidable \cite{Norman}. This leads to the notorious magnetic disorder effect, and structural imperfections of its kagom\'{e} lattice. They can cause spurious signals in experiments mimicking QSL features~\cite{Sasha}, and can also diminish the delicate sought-after signatures of QSL states~\cite{Yamashita}.
The difficulties of extracting pristine information from herbertsmithite experiments lead to heated debates on many key issues, halting feedbacks to theoretic considerations of the KHA model~\cite{WenXG,PAL,HeYC,Iqbal,MeiJW,JiangHC}.
What renders the situation even more intricate is that theoretical works found different possible ground states of the KHA model which are very close in energy \cite{Norman,JPCS}.
Needless to say, searching for another representative material of the KHA model is of paramount demand.

For these reasons, the importance of identification of the recently synthesized bromine compound YCu$_3$(OH)$_{6.5}$Br$_{2.5}$ (YCOB) as a new potential KHA material is obvious~\cite{ChenXH,LiYS,LiSL}.
As illustrated in Fig.~1, spin-1/2 Cu$^{2+}$ ions fully occupy the globally regular kagom\'{e} sites, free from site mixing with other nonmagnetic ions.
The magnetic system of YCOB is dominated by large nearest neighbor coupling of an averaged value $J_1$ $\sim$ 60~K, more than one order of magnitude larger than the further-neighbor and interlayer couplings~\cite{LiYS}.
A certain exchange bond randomness is, however, to be expected because the distribution of OH$^-$/Br$^-$ is not symmetric in between the Cu kagom\'{e} layers, pushing up to~70\% of Y$^{3+}$ ions out of their ideal position~\cite{LiYS}.
As a candidate for the KHA, no magnetic order was detected in YCOB by preliminary thermodynamic studies down to 50 mK ($< J_1$/1000) \cite{ChenXH,LiYS,LiSL}.
In brief, YCOB appears to be the most promising realization of the kagom\'{e} QSL without evident orphan spins so far.

\begin{figure}
\includegraphics[clip,width=0.4\textwidth]{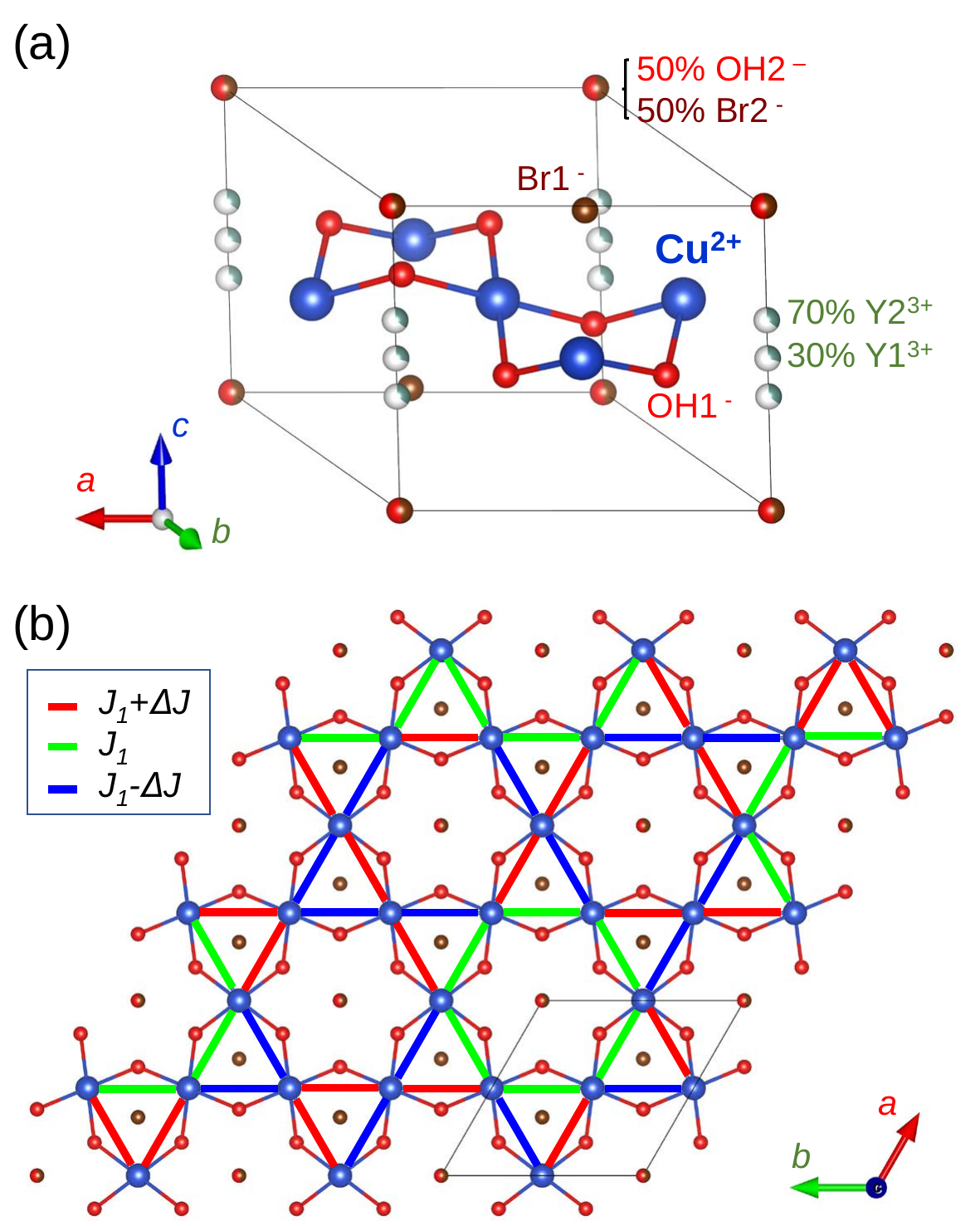}
\caption{
(a) Lattice structure of YCOB. Although the antisite mixing of OH2$^-$/Br2$^-$ pushes 70\% of Y$^{3+}$ away from its ideal position and imposes quenched exchange randomness on the kagom\'{e} bonds, as illustrated in (b), the space group of $P\bar{3}m1$ and thus the globally regular kagom\'{e} lattice of Cu$^{2+}$ remains intact~\cite{LiYS}. The black lines show the unit cell.
}
\end{figure}

In this letter, low-temperature thermal conductivity ($\kappa$) measurements were performed on three single crystals of YCOB down to 70 mK. We firmly conclude that there is no residual linear term in the $T\rightarrow$ 0 extrapolation of the $\kappa/T$ curve ($\kappa_0/T$), i.e. there are no gapless fermionic excitations, which for example, are expected to emerge from a spinon Fermi surface.
On the other hand, a clear deviation off a standard phononic power-law $\kappa/T$ curve is observed at a quite low temperature $T$*, implying that phonon-spin scattering persists down to mK range. Additionally, a clear magnetic field dependence of $\kappa$ is uncovered, revealing the field impact on the spin excitation spectrum of YCOB. Our analysis yields a spin gap $\Delta/k_B \approx 1.4$~K at zero magnetic field, which reduces upon application of large magnetic fields.

Single crystals of YCOB were grown by the hydrothermal technique~\cite{ChenXH}, followed by a recrystallization process to improve the crystal quality \cite{LiYS}. Three samples of roughly rectangular shape of 1.8$\times$0.24$\times$0.18 (Sample\#1), 0.85$\times$0.21$\times$0.053 (Sample\#2), and 2.2$\times$0.32$\times$0.24 (Sample\#3) mm$^3$ were used for the heat transport measurements. Four silver wires were attached to the surfaces of the samples with silver paint, to let the heat flow in the $ab$-plane. Thermal conductivity was measured in a dilution refrigerator using a standard four-wire steady-state method, with two RuO$_2$ chip thermometers, which were $in$-$situ$ calibrated against a reference RuO$_2$. Magnetic field was applied out of the $ab$-plane.

\begin{figure}
\includegraphics[clip,width=0.46\textwidth]{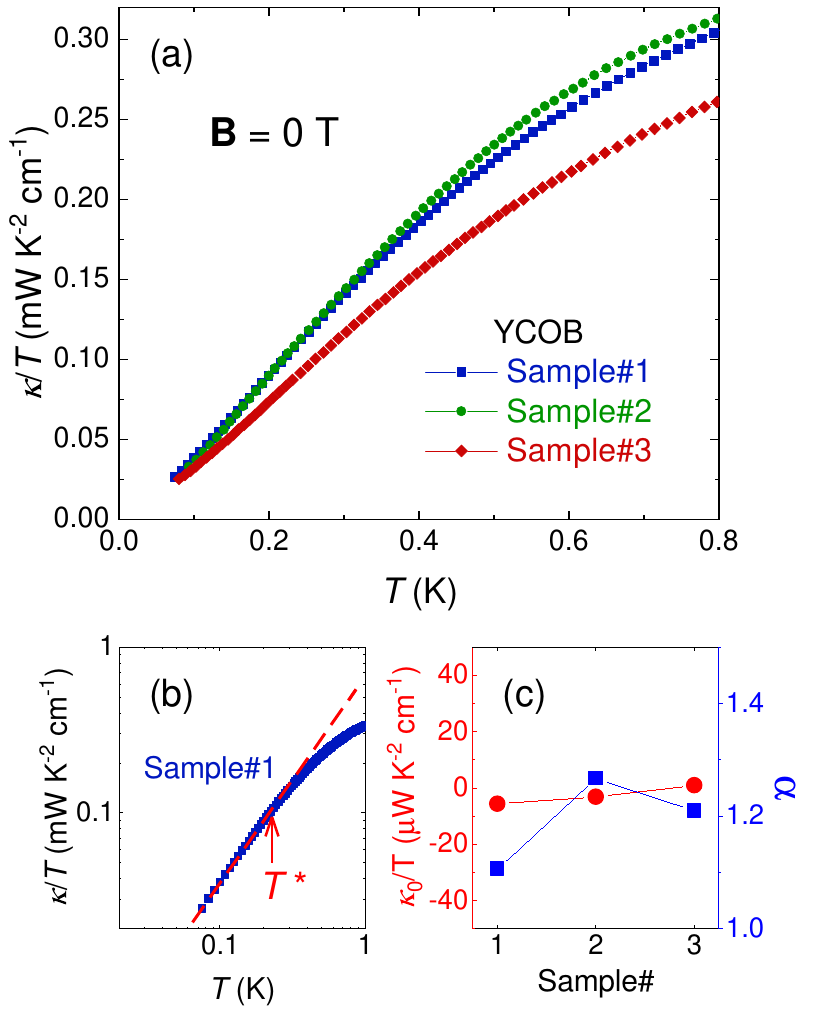}
\caption{ Low-temperature thermal conductivity of YCOB crystals without magnetic field.
(a) $\kappa/T$($T$) curves of three single crystals.
(b) $\kappa/T$($T$) curve of Sample\#1 as a representative case, plot in the log-log scale. The dashed red line highlights its linear dependence at the low temperature limit. The data deviate downward from this linear behavior above a temperature $T$*~$\sim 230$ mK marked by the arrow.
(c) The fit parameters residual liner term ($\kappa_0/T$, left axis) and exponent ($\alpha$, right axis) of all three samples.
}
\end{figure}

The thermal conductivity of the three YCOB samples without magnetic field, plotted as $\kappa/T$ $vs$ $T$, is shown in Fig. 2(a). They are of very similar type. The minor difference of absolute values of the calculated $\kappa/T$ can be attributed to the uncertainty in the measurement of the sample geometry.
The first key message of our work is the clear trend that all $\kappa/T$ curves extrapolate towards the origin of coordinates. This fact precludes itinerant gapless quasiparticles of a Fermi surface contributing to the thermal transport. Otherwise, these should yield a sizeable residual term $\kappa_0/T$, like the electronic heat conductivity in metals~\cite{dmit}. The data suggest that phonons are the only heat carrier in YCOB, and fractionalized fermionic excitations, if they exist, have no Fermi surface or are localized.

Indeed, at $T\rightarrow0$ all curves obey a power law $\kappa/T\propto T^{\alpha}$ with $\alpha$ between 1.1 and 1.3, see Fig. 2(b) and (c), see below for details. This represents the expectation of $1<\alpha\leq2$ for standard ballistic phonon transport in the low-temperature limit, where the mean free path of the phonons is limited by scattering off the sample surface.
Note, that on naturally grown crystal surfaces, like is the case here, the phonon reflection at the sample surface normally is not perfectly diffuse, but becomes increasingly specular as the mean phonon wave length grows upon lowering the temperature. Thereby, the phonon means free path also effectively grows, and one expects $1<\alpha<2$ \cite{Pohl,Thacher,Pohl0,LiSY}.
This indeed is the case in our samples according to a fit
\begin{equation}
\frac{\kappa}{T} = \frac{\kappa_0}{T} + bT^\alpha~,
\end{equation}
which yields a vanishing small $\kappa_0/T$ and a slightly sample dependent $\alpha$ in the mentioned range~\cite{SM}, as summarized in Fig. 2(c).

However, for all samples the phononic low-temperature power law is violated above a critical temperature $T$* where $\kappa/T$ deviates downward from it, see Fig.~2(b) for the representative case of Sample\#1 \cite{SM}.
The reduction of $\kappa/T$ clearly indicates the breakdown of the ballistic phonon heat transport and implies additional scattering of the phonons, the only source of which can be phonon-spin scattering.
Actually, recent low-temperature heat transport studies on several quantum magnets revealed this type of scattering being important whenever magnetic fluctuations prevail \cite{SunXF,XuY,Matthias,YuYJ}.
Hence, since below $T$* such scattering apparently freezes out in YCOB, without further analysis, our data imply an energy gap of the spin fluctuations.

\begin{figure}
\includegraphics[clip,width=0.45\textwidth]{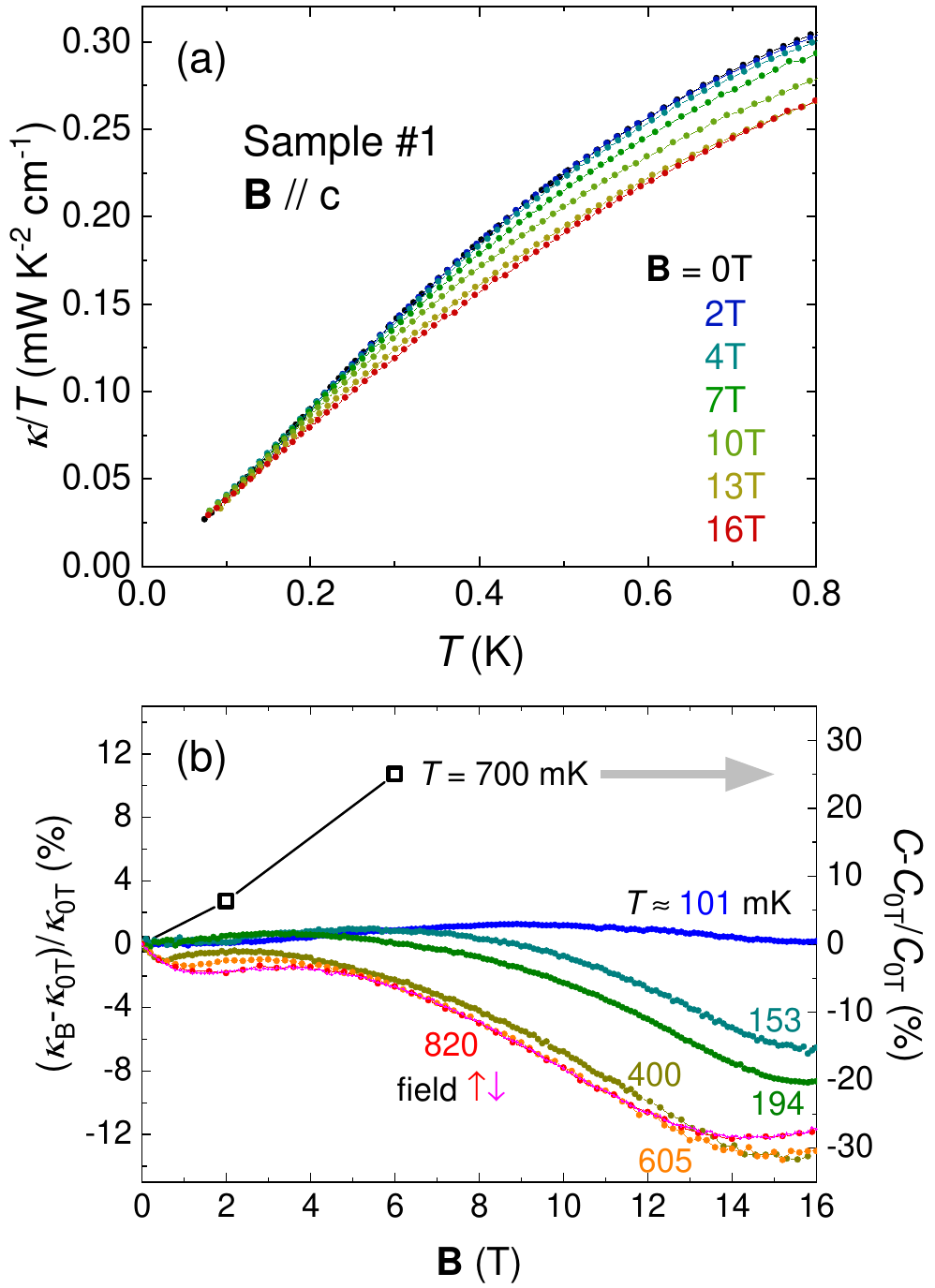}
\caption{ Field impact on the thermal conductivity of YCOB.
(a) The $\kappa/T$ curves at different fields. Increasing magnetic field suppresses the $\kappa/T$ curve monotonically, and this effect is discernible at lower temperatures in higher fields.
(b) The closed color circles (left axis) represent the relative change of $\kappa$ according to the field change at different temperatures. For the $T =$ 820 mK curve, data collected with field ramped up (larger red circles) and down (smaller magenta circles) match perfectly, excluding hysteresis effects. The open squares (right axis) show the relative field change of the magnetic specific heat, adapted from Ref. \cite{LiYS}.}
\end{figure}

To gain more information about the ground state properties of YCOB, $\kappa$($T$) was measured in various magnetic fields up to 16 T.
The full set of results of Sample\#1 are shown in Fig. 3 as a representative case. Apparently, all three samples share very similar behavior~\cite{SM}.
$\kappa$($T$) curves get further suppressed with increasing field, and $T$* is concomitantly pushed to lower temperatures.
This sensitivity of $\kappa$ implies the field tunability of the magnetic excitations spectrum, underpinning the low energy nature of these excitations.

$\kappa$($H$) isotherms can provide information of the field impact on thermal transport from another perspective. As displayed in Fig. 3(b), a field of 16 T can reduce $\kappa$ by about 12.5\% at higher temperatures \cite{SM}. The field effect is less significant at lower temperatures, and is nearly indistinguishable at the lowest temperature investigated (101mK, below $T$* at 16 T).
It is worthwhile to notice that the magnetic specific heat $C_{mag}$ of YCOB increases in field \cite{LiYS,LiSL}.
Such anticorrelated $\kappa$ and $C_{mag}$ is hard to reconcile if they are contributed directly from the same (quasi)particles.
Normally, for any entropy carrier, its contribution to $\kappa$ is proportional to $C$ multiplied by its velocity $v$ and its mean free path $l$, $\kappa_i \propto C_iv_il_i$. The total thermal conductivity and specific heat are the sum of all components, $\kappa$ = $\sum\limits_i \kappa_i$ and $C$ = $\sum\limits_i C_i$. Hence, the opposing trends of $\kappa$ and $C$ in field can only be reasonably explained by accepting that the field sensitive quasiparticle contributes differently to these two quantities.
More specifically, since field enhanced $C_{mag}$ means a proliferation of magnetic excitations, the observed reduction of $\kappa$ corroborates our above conclusion that a direct contribution of magnetic quasiparticles to the heat transport is absent, unless the heat current relaxation rate overcompensates the increase in specific heat in an unusual way.
In contrast, the magnetic quasiparticles apparently have rather a clear detrimental effect on the omnipresent phononic heat transport channel. I.e., significant phonon-spin scattering must be active which reduces the phonon mean free path.
Hence, the field induced suppression $\kappa$ further corroborates our scattered phononic heat transport scenario proposed earlier to explain the zero-field $\kappa$($T$) behavior.

Having established that magnetic fluctuations do not directly contribute to the heat transport in YCOB but have a significant impact on the phononic transport beyond $T$* through scattering, we now turn to a more detailed analysis. According to Matthiessen's rule, an additional magnetic scattering mechanism affects the total scattering time ($\tau_{total}$) in the form of
\begin{equation}
\tau_{total}^{-1} = \tau_p^{-1} + \tau_{ps}^{-1}~,
\end{equation}
where $\tau_p$ and $\tau_{ps}$ describe the intrinsic phononic and the additional phonon-spin relaxation times, respectively. A $\kappa$($T$) curve free from phonon-spin scattering, $\kappa_{fit}$($T$), is estimated by the power-law fit to the zero-field $\kappa$($T$) curve below $T^*$ and its extension to the whole temperature range (dashed line in Fig. 2(b)).
We seek to find a direct measure of the phonon-spin scattering rate $\tau_{ps}^{-1}$. Our data provide access to it via the normalized reduction of the hypothetical intrinsic phononic heat conductivity $\kappa_{fit}$ with respect to the measured $\kappa$~\cite{SM}:
\begin{equation}
\frac{\kappa_{fit} - \kappa}{\kappa} = \frac{\tau_{p}}{\tau_{ps}} \propto \frac{1}{\tau_{ps}}~,
\end{equation}
where the last proportionality is roughly valid due to the fact that the intrinsic relaxation time is only weakly temperature dependent in the low-temperature ballistic limit considered here.

\begin{figure}
\includegraphics[clip,width=0.49\textwidth]{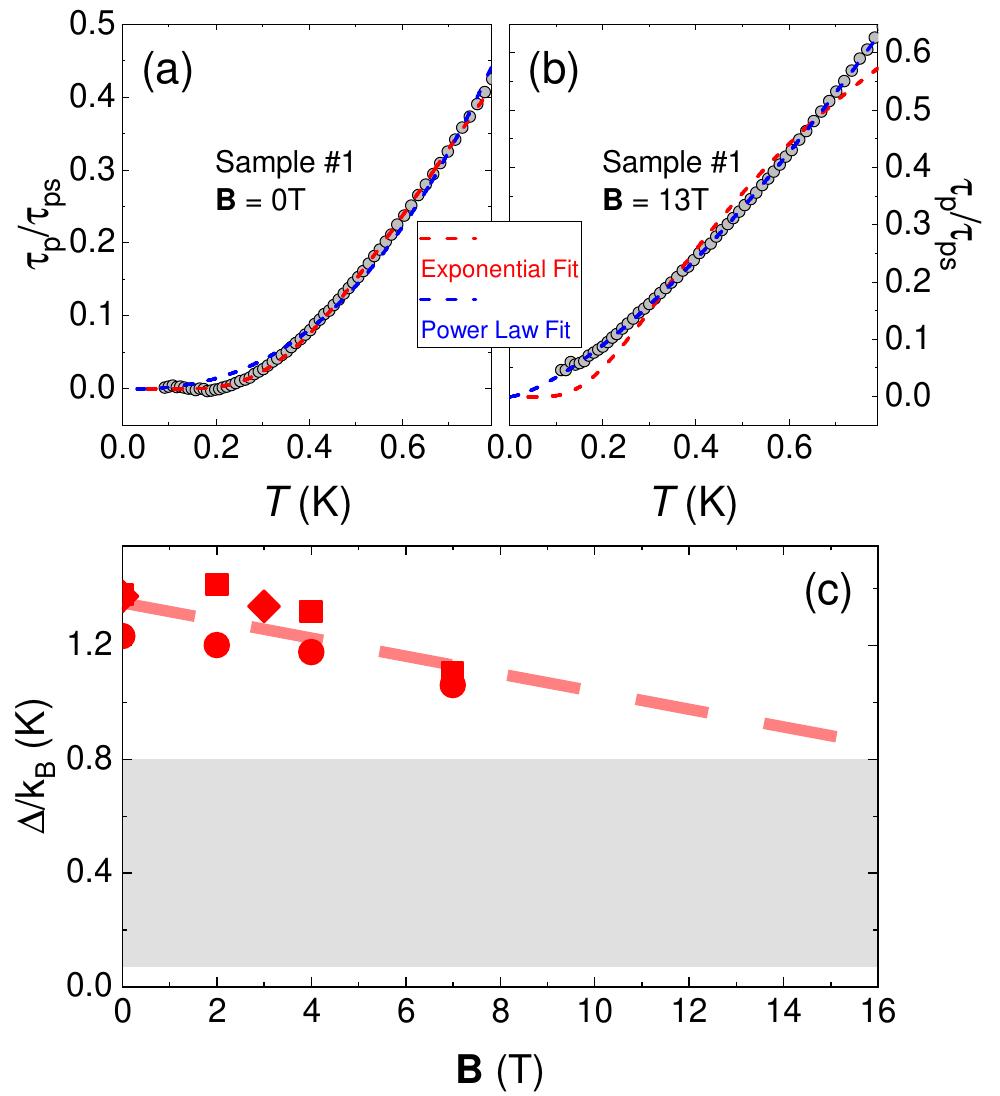}
\caption{Exponential and power-law fits to the temperature dependence of $\tau_p$/$\tau_{ps}$ (see text) for one sample at (a) zero field, and (b) high field. (c) The gap size estimated from the acceptable exponential fits at different fields of Sample\#1 ($\blacksquare$), Sample\#2 ($\bullet$), and Sample\#3 ($\blacklozenge$)~\cite{SM}. The gray region marks the temperature range over which the fit were performed. The dashed line is a guide to the eye (see text).}
\end{figure}

As presented in Fig. 4(a), an exponential function
\begin{equation}
\frac{\tau_{p}}{\tau_{ps}} = ne^{-\Delta/(k_BT)}~,
\end{equation}
with $n$ a proportionality constant, almost perfectly fits the zero-field result.
Such thermally activated enhancement of the phonon-spin relaxation rate is only compatible with theoretical results which predict the ground state of KHA model being a gapped $\mathbb{Z}_2$ state~\cite{PAL,MeiJW,JiangHC}, if in addition the parton excitations which are necessarily present in such models and moreover provide a natural source for such activated phonon relaxation are localized.
This seems likely, since the quenched exchange randomness due to the antisite mixing (see Fig.~1) implies off-diagonal disorder in a two-dimensional parton mean-field theory and moreover the kagom\'{e} lattice is non-bipartite \cite{Abrahams,Xiong,Note,Note2}.
The estimated excitation gap $\Delta/k_B \sim 1.4$~K is quite small compared to the average $J_1 \sim 60 K$, and is in good agreement with that obtained from fitting the specific heat according to the gapped $\mathbb{Z}_2$ scenario, $\Delta$ $\sim$ 0.025$J_1$ ~\cite{LiYS}.

In the $\mathbb{Z}_2$ scenario for the ground state, the observed enhanced suppression upon application of magnetic fields implies a diminuition of the excitation gap. Indeed, for all samples Eq. 4 yields good fits at moderate magnetic fields~\cite{SM}, where the extracted gap $\Delta(B)$ decreases with increasing field.
The simplest way a magnetic field could affect a triplet excitation gap is due to Zeeman energy which is linear in field, as has previously been suggested to be relevant in herbertsmithite~\cite{FuMX}, see Fig.~4(c).
However, at fields larger than about 7~T, much lower than the extrapolated gap closing field, a simple exponential temperature dependence already fails to describe the data, as exemplified by Fig. 4 (b). Instead, a power-law fit
\begin{equation}
\frac{\tau_{p}}{\tau_{ps}} = AT^\alpha~,
\end{equation}
which corresponds to a gapless excitation spectrum, much better describes the data above 10~T~\cite{SM}.

At first glance, this seems to suggest a gapped to gapless quantum phase transition somewhere between 7~T and 10~T. We point out that in this case, a reduction of the phonon mean free path should also occur at lowest temperatures and at fields larger than 10~T. However, as can be inferred from Fig.~3(b), at the lowest temperature measure ($T_{min} = 101$~mK), within resolution, a significant magneto-heat conductivity is barely resolvable, which is compatible with $\Delta$ being finite and larger than $k_BT_{min}$ even at $B=16$~T. Hence, the good agreement of the high field data with a power law temperature dependence should not be interpreted as a complete closing of a gap. Instead one can expect it to arise due to the fact that at $B\gtrsim7$~T the thermal energies $k_BT$ in the low-temperature measurement range are already of similar order than $\Delta$. In this situation, not only the lowest lying magnetic excitations contribute to the scattering of phonons but a considerable fraction of the magnetic excitation spectrum.
We point out that the same argument also explains why the previous analysis of the specific heat data both lead to the conclusion of gapless excitations in YCOB at zero field~\cite{LiYS,LiSL}.
If one takes into account that these experiments were carried out at much higher temperatures than our heat transport study, the reported data are well compatible with a gapped excitation spectrum~\cite{LiYS,LiSL}.

Before we conclude, we point out that the zero field data may leave some room for an alternative scenario of magnetic quasiparticles directly contributing to the heat transport. In this case the observed low-$T$ power-law in $\kappa/T$ would imply gapless magnetic excitations.
In finite field these excitations must remain gapless because the power law is preserved at lowest $T$. While this is difficult to reconcile with the concomitant field-induced suppression of $\kappa/T$ at higher $T$, the scenario of gapless heat carrying quasiparticles still cannot be completely excluded.
In order to fully rule it out, an out-of plane heat transport ($\kappa_c$) experiment for disentangling magnetic and phononic contributions to $\kappa$ would  be required \cite{Hess}. Such an experiment is, however, hardly conceivable since the available crystals are extremely thin platelets with large surfaces parallel to the kagom\'{e} layers.

In summary, by measuring the low-temperature thermal conductivity of YCOB, we managed to probe the excitations in this realization of a KHA model. The absence of direct magnetic contribution to $\kappa$ firmly exclude itinerant spin excitation in this compound on top of its ground state.
Furthermore, utilizing scattered phononic thermal conductivity as a probe, we demonstrated that the magnetic excitation spectrum  in YCOB is gapped, consistent with a $\mathbb{Z}_2$ spin liquid ground state. The gap size is reduced by applying magnetic field, but remains finite up to $B = 16$~T.
Our study highlights YCOB as a prime kagom\'{e} QSL candidate free from severe disorder problems, an ideal testbed for various QSL theories.

This work has been supported by the Deutsche Forschungsgemeinschaft (DFG) through  SFB 1143 (Project-id No. 247310070).
This project has received funding from the European Research Council (ERC) under the European Union's-Horizon 2020 research and innovation programme (grant agreement No. 647276-MARS-ERC-2014-CoG). The work at HUST was supported by the Fundamental Research Funds for the Central Universities, HUST: 2020kfyXJJS054.


\begin{thebibliography}{99}
\bibitem{Note} We assume that the intrinsic bond randomness does not lead to in-gap states.
\end{thebibliography}
\end{document}